\documentstyle[aps,prl,twocolumn]{revtex}

\begin{document}
\draft

\title{Quantum entanglement and information processing \\
via excitons in optically-driven quantum dots}

\author{John H. Reina$^{1\aleph}$,
Luis Quiroga$^{2\dagger},\;$and Neil F. Johnson$^{1\ddagger}$}

\address{$^{1}$Physics Department, Clarendon Laboratory, Oxford
University, Oxford, OX1 3PU, England}

\address
{$^{2}$Departamento de F\'{\i}sica, Universidad de Los Andes, Santaf\'{e}
de Bogot\'{a}, A.A. 4976, Colombia }

\twocolumn[\hsize\textwidth\columnwidth\hsize\csname
@twocolumnfalse\endcsname
\maketitle

\begin{abstract}We show how optically-driven coupled quantum dots can be used
to prepare maximally entangled Bell and Greenberger-Horne-Zeilinger states.
Manipulation of the strength and duration of the selective light-pulses needed
for producing these highly entangled states provides us with crucial elements
for the processing of solid-state based quantum information. Theoretical
predictions suggest that several hundred single quantum bit rotations and
Controlled-Not gates could be performed before decoherence of the excitonic
states takes place.
\end{abstract}


\vskip2pc]
\narrowtext

\section{Introduction}

Quantum computation, quantum communication, quantum cryptography and
quantum teleportation
\cite{Feynman,Deutsch,computation,Cryptography,Bennett1} are some of the
most exciting applications of the fundamental principles of quantum
theory. Since the seminal idea of Feynman in 1982 \cite{Feynman} and the work 
of Deutsch in 1985 \cite{Deutsch}, both pure and applied research in the field of 
quantum information processing has blossomed. In 1994, Shor \cite{Shor1} 
opened the way to new fast quantum searching algorithms:
he discovered that a quantum computer can factorize large integers. Two years
later arrived the proof that quantum error-correcting codes exist
\cite{Shor2}. Up until now, such quantum mechanical computers have been
proposed in terms of trapped ions and atoms \cite{Trapped}, cavity quantum
electrodynamics (QED) \cite {Cavity QED}, nuclear magnetic resonance
\cite{NMR}, Josephson junctions  \cite{Superconductor} and semiconductor
nanostructures
\cite{Semiconductors} schemes. All of the above\mbox{ proposals\/} have decoherence and
operational errors as the main obstacles for their experimental \mbox{ realization\/},
which pose much stronger problems here than in classical computers.

There is much current excitement about the\mbox{ possibility\/} of using solid-state
based devices for the achievement of quantum computation tasks. In particular, 
semiconductor nanostructure fabrication technology is well developed and hence
offers us a wide and promising arena for the challenging project of building quantum
information processors. Because of their quantum mechanical nature and their potential
scalability properties, semiconductor quantum dots
(QDs) are very promising candidates for the implementation
of quantum computing processes. Several solid state design schemes for quantum computation
have been proposed to date: Kane \cite{Semiconductors} has proposed a scheme which
encodes information onto the nuclear spins of donor atoms in doped silicon
electronic devices where externally applied electric fields are used to
perform logical operations on individual spins. Loss and DiVincenzo \cite{Semiconductors} 
have presented a scheme based on electron spin effects, in which coupled quantum dots are 
used as a quantum gate. This scheme is based
on the fact that the electron spins on the dots have an exchange interaction
$J$ which changes sign with increasing external magnetic field. Possible
quantum gate implementations have also been proposed by Barenco {\it et al.}
\cite{Semiconductors} by\mbox{ considering\/} electronic charge effects in coupled QDs,
however this scheme has as the main disadvantage rapid phonon decoherence, as
compared with the above proposals. More recently, Imamoglu {\it et al.}
\cite{Semiconductors} have\mbox{ considered} a quantum computer model based on both
electron spins and cavity QED which is capable of realizing controlled
interactions between two distant QD spins. In their model, the effective
long-range interaction is mediated by the vacuum field of a high finesse
microcavity, and single quantum bit (qubit) rotations and Controlled-Not
(CNOT) operations are realized using electron-hole Raman transitions
induced by classical laser fields and the cavity mode. Vrijen
{\it et al.} \cite {Semiconductors} considered electron spin resonance
transistors in Silicon-Germanium heterostructures: one and two qubit
operations are performed by applying a gate bias.

In this paper we focus on optically-driven coupled QDs 
in order to obtain highly entangled states of excitons and
hence provide a mechanism for processing quantum information over a reasonable
parameter window, before decoherence of the excitonic states takes place.
In the physical implementation of the quantum entanglement scheme proposed
here, we exploit recent experimental results involving {\it coherent optical
control} of excitons in single quantum dots on the nanometer and femtosecond
scales \cite{Bonadeo1,Bonadeo2,ChavezPirson}. The amazing degree of control
of the quantum states of these individual \mbox{\ ``artificial atoms''\/} \cite
{Bonadeo1} due to the manipulation of the confined state wave function of a
single dot is an exciting and promising development. As one exciton can be trapped in
the dot, we have the direct possibility to use QDs as elements with {\it
quantum memory} capacity in quantum computation operations, through a precise
and controlled excitation of the system. As demonstrated in \cite
{Bonadeo1}, it is\ possible to excite and probe {\it only one} individual QD
within a broad distribution of dots{\it , }with the important result that the\mbox{ dephasing} 
time is much longer in a single dot ($40$ps) than in the bulk
semiconductors ($<1$ps) studied before. Hence, new
experimental work means that much longer intrinsic
coherence times are currently available, a fact of fundamental
importance when looking towards the practical implementation of
single qubit rotations and quantum CNOT gates which are crucial elements for
carrying out quantum information processing tasks. 

The outline of this paper is as follows: Section II gives a detailed
description of the coupling of qubits (i.e. QDs) to pulses of light and the
generation of maximally entangled Bell (or EPR) \cite{Bell,Einstein} and
Greenberger-Horne-Zeilinger (GHZ) \cite{GHZ} states$\;$in systems comprising
two and three quantum dots, respectively. The discussion of the results and
experimental considerations are presented in Section III. In Section IV
we review the elements required for performing quantum computation
tasks and the links with the solid state set-up proposed here. Conclusions are
given in Section V.

\section{\mbox{\ generation of
maximally entangled\/} states in optically-driven Quantum Dots}

Quantum entanglement is of fundamental interest since many of the most basic
aspects of quantum\mbox{ theory\/} require its successful generation and
manipulation. In particular Bell and GHZ maximally entangled states are the
starting point for fundamental discussions such as the violation of Bell's
inequalities \cite{Aspect} and the non-locality problem \cite {Einstein}, as
well as for teleportation \cite{Bennett1} and quantum cryptography
\cite{Cryptography}. Here we will show how to generate maximally entangled
states of two and three qubits of the form $\left| \Psi _{Bell}\right\rangle =
{\textstyle{1 \over \sqrt{2}}}
(\left| 00\right\rangle +e^{i\varphi }\left| 11\right\rangle )\;$and $\left|
\Psi _{GHZ}\right\rangle =
{\textstyle{1 \over \sqrt{2}}}
(\left| 000\right\rangle +e^{i\varphi }\left| 111\right\rangle )$,$\;$for
arbitrary values of the phase factor $\varphi $ \cite{Note0}, using a semiconductor 
nanostructure set-up. We consider a
system of $N$ identical and equispaced QDs, containing no net charge, which
are radiated by long-wavelength classical light. Hence formation of single
excitons within the individual QDs and their inter-dot transfer can be
described in the frame of the {\it rotating wave approximation} (RWA), by
the Hamiltonian \cite{Quiroga,Note1} ($\hbar=1$): 

\begin{eqnarray}
\nonumber
H(t)=
{\textstyle{\epsilon  \over 2}}
\mathop{\displaystyle\sum}
\limits_{p=1}^{N}\{c_{p}^{\dagger {}}c_{p}-h_{p}h_{p}^{\dagger {}}\}+
{\textstyle{W \over 2}}
\mathop{\displaystyle\sum}
\limits_{p,p^{\prime }=1}^{N}\{c_{p}^{\dagger {}}h_{p^{\prime }}c_{p^{\prime
}}h_{p}^{\dagger {}}+
\\
h_{p}c_{p^{\prime }}^{\dagger {}}h_{p^{\prime
}}^{\dagger {}}c_{p}\}+\xi (t)
\mathop{\displaystyle\sum}
\limits_{p=1}^{N}c_{p}^{\dagger {}}h_{p}^{\dagger {}}+\xi ^{\ast }(t)
\mathop{\displaystyle\sum}
\limits_{p=1}^{N}h_{p}c_{p}\text{,} 
\end{eqnarray}
where $c_{p}^{\dagger {}}\;(h_{p}^{\dagger {}})\;$is the electron (hole)
creation operator in the $p$th quantum dot, $\epsilon $ is the QD band gap, $W$ the interdot interaction, 
and $\xi (t)$ the laser pulse shape. The
operators involved in Eq. (1) obey the anticommutation rules $\left\{ c_{p^{\prime
}},c_{p}^{\dagger {}}\right\} =\left\{ h_{p^{\prime }},h_{p}^{\dagger
{}}\right\} =\delta _{pp^{\prime }}$. By introducing the new operators 
\begin{eqnarray}
\nonumber
J_{+}=
\mathop{\displaystyle\sum}
\limits_{p=1}^{N}c_{p}^{\dagger {}}h_{p}^{\dagger {}}\text{,}%
\;J_{-}=
\mathop{\displaystyle\sum}
\limits_{p=1}^{N}h_{p}c_{p}\text{%
,\ }
\\
J_{Z}=
{\textstyle{1 \over 2}}
\mathop{\displaystyle\sum}
\limits_{p=1}^{N}\{c_{p}^{\dagger {}}c_{p}^{{}}-h_{p}h_{p}^{\dagger {}}\}%
\text{, } 
\end{eqnarray}
Eq. (1) adopts the form 
\begin{equation}
H(t)=H_{0}+H_{L}(t)\text{,}
\end{equation}
where 
\begin{equation}
H_{0}=\epsilon J_{Z}+W(J^{2}-J_{{\small Z}}^{^{2}})\text{,}\;\; H_{L}(t)=\xi (t)J_{+}+\xi ^{\ast }(t)J_{-}\text{,}
\end{equation}
with $J^{2}\equiv
{\textstyle{1 \over 2}}
[J_{+}J_{-}+J_{-}J_{+}]+J_{Z}^{^{2}}$.$\;$ The \mbox{\ $J_{i}-$operators o-\/}\\
bey the usual angular momentum commutation relations $\left[J_{Z},J_{\pm } \right] =\pm
J_{\pm },\;\left[ J_{+},J_{-}\right] =2 J_{Z},\;
$and$
\; [J^{2},J_{+}]=[J^{2},J_{-}]=[J^{2},J_{Z}]=0.$
In going from Eq. (1) to Eq. (3) we have switched from a
dot-selective (index $p$) to a non-selective description. Since the dots are
equidistant from each other, this description is appropriate for up to 4
dots (placed at the corners of a tetrahedron). Hence the quantum dynamical
problem associated with the time evolution of any initial state under the
action of
$H(t)$ is described by:
\begin{equation}
i\partial _{t}\left| \Psi (t)\right\rangle _{{\small S}}=
\{H_{0}+H_{L}(t)\}\left| \Psi (t)\right\rangle _{{\small S}}\text{,}
\end{equation}
where the subscript $S$ indicates Schr\"{o}dinger picture. We consider the
laser pulse shape $\xi (t)=Ae^{-i\omega t}$, where $A$ gives the
electron-photon coupling and the incident electric field strength$. $ We also
introduce the unitary transformation $\Lambda (t)=e^{-i\omega J_{Z}t}$, whose
application in the Schr\"{o}dinger picture leads us from the {\it
laboratory frame} (LF) to the {\it rotating frame} (RF) using the rule
$\left| \Psi (t)\right\rangle _{\Lambda }=\Lambda ^{\dagger }(t)\left| \Psi
(t)\right\rangle _{S}$. Hence Eq. (5) may be\mbox{ rewritten} as: 
\begin{equation}
i\partial _{t}\left| \Psi (t)\right\rangle _{\Lambda }=H^{\prime
}\left| \Psi (t)\right\rangle _{\Lambda }\;\text{,}
\end{equation}
with 
\begin{equation}
H^{\prime }=\Delta _{\omega }J_{Z}+W(J^{2}-J_{{\small Z}}^{2})+AJ_{+}+A^{%
\ast }J_{-}\;\text{.}
\end{equation}
Here $\Delta _{\omega }\equiv\epsilon-\omega$ is the detuning
parameter. We note the importance of the $\Lambda - $transformation: the new
Hamiltonian $H^{\prime }$ is time-independent.$\;$From a practical point of
view, parameters $A\;$and $\Delta _{\omega }$\ are adjustable in the
experiment to give control over the system of QDs (or qubits). Since $J^{2}$ commutes
with the\mbox{ operators\/} $J_{\pm }$, $H^{\prime }$ may be \mbox{\ diagonalized\/}
separately in each one of these $J-$subspaces. Consider the $\{J,q\}$ subspace
spanned by $\left| M\right\rangle \equiv \left| J,M;q\right\rangle $: the only
possible values for $J$ are $\frac{N}{2},\frac{N}{2}-1,...,\frac{1}{2}\;$or$
\;0$,$\;$and\ for\ each $J-$fixed value, we have $2J+1$ different values for 
$M$, which are given by$\;M=-\frac{N}{2},-\frac{N}{2}+1,...,\frac{N}{2}-1,
\frac{N}{2}.$  We introduce the label $q\;$to further distinguish the states:
$\ q=1,2,...,D_{J}$, where the multiplicity $D_{J}$, i.e. the number of states
having angular momentum $J$ and $M=J$, is given by $D_{J}=\frac{2J+1}{
J+\frac{N}{2}+1}{N \choose \frac{N}{2}+J}$. The\mbox{ product\/} states
$\prod_{k=1}^{N}\left| m_{k}\right\rangle
\equiv\left| m_{1},...,m_{N}\right\rangle, $ with $J_{Z}=
\mathop{\textstyle\sum}
_{k}m_{k}$ form a $2^{N}-$dimensional basis which span the Hilbert space$\
SU(2)^{\otimes N}$. In this basis, the $2^{N}$ eigenvalues of $H^{\prime }$\
are obtained by diagonalizing the Hamiltonian matrix of elements $%
\left\langle J,M,q\right| H^{\prime }\left| J^{\prime },M^{\prime
},q^{\prime }\right\rangle $. We get the non-zero elements as follows: 

\begin{eqnarray}
\nonumber
\left\langle M\right| H^{\prime }\left| M^{\prime }\right\rangle = 
\left( \Delta _{\omega }M+W\left[
J(J+1)-M^{2}\right] \right) \delta _{{\small M,M}^{\prime }}+  \nonumber \\
&& \hspace{-6.4cm}
A\sqrt{J(J+1)-M^{\prime }(M^{\prime }+1)}\;\delta _{{\small M,M}^{\prime }%
{\small +1}}+ \nonumber \\ 
&& \hspace{-6.4cm}
A^{\ast }\sqrt{J(J+1)-M^{\prime }(M^{\prime }-1)}\;\delta _{%
{\small M,M}^{\prime }{\small -1}}\text{.}\end{eqnarray}
The matrix elements given in Eq. (8) provide us with the general rule for
any number of QDs. Since the right side of this equation does not
depend on $q$ we only need to diagonalize a square matrix of side $2J+1$ for
each $J$. \mbox{ Every\/} eigenvalue so obtained occurs $D_{J}$ times in the entire
spectrum. Next, we show that solving the eigenfunction problem
associated with Eq. (8) leads to the generation of highly $N-$entangled
states of excitons in QDs.

\subsection{Coupling of $\bf N=2$ QDs and generation of Bell states}

In this section we describe the procedure for the \mbox{ generation\/} of entangled
Bell states $\left| \Psi _{Bell}\right\rangle = 
{\textstyle{1 \over \sqrt{2}}}
(\left| 00\right\rangle +e^{i\varphi }\left| 11\right\rangle )\;$ for
arbitrary values of the phase factor $\varphi $. Here $0$ ($1$) denotes a
zero-exciton (single exciton) QD, and the direct product of the quantum
states $\left| jk\right\rangle \equiv \left| j\right\rangle \otimes \left|
k\right\rangle $ form a four-dimensional basis in the Hilbert space $%
SU(2)\otimes SU(2)$. In the $J=1$ subspace \cite{Note2},$\;M\equiv \{-1,0,1\}
$.$\;$We define $\left| M_{1}\right\rangle \equiv \left|
J=1,M=-1\right\rangle \equiv \left| 0\right\rangle $, $\left|
M_{2}\right\rangle \equiv \mbox{\ $\left| J=1,M=0\right\rangle \equiv \left|
1\right\rangle, $\/} $ and $\left| M_{3}\right\rangle \equiv \left|
J=1,M=1\right\rangle \equiv \left| 2\right\rangle $, as the vacuum of
excitons$,$ the single-exciton state and the biexciton
state respectively. In the absence of light, we have:
\begin{equation}
E(J,M)=\Delta _{\omega }M+W[J(J+1)-M^{2}]\text{,}  
\end{equation}
so the energy levels of the system are $E_{0}\equiv
E(1,-1)=W-\Delta _{\omega }$, $E_{1}\equiv E(1,0)=2W$, and $E_{2}\equiv
E(1,1)=W+\Delta _{\omega }$. Note that $E_{2,0}\equiv E_{2}-E_{0}=2\Delta
_{\omega }$ is unaffected by the interdot interaction strength $W$. Next,
consider the action of the radiation pulse $\xi (t)$ over this pair of
qubits; in the $J=1$ subspace the Hamiltonian adopts the simple form: 
\begin{equation}
\widehat{H^{^{\prime }}}=\left( 
\begin{array}{ccc}
W-\Delta _{\omega }\; & \;\sqrt{2}A^{\ast } & 0 \\ 
\sqrt{2}A & 2W & \sqrt{2}A^{\ast } \\ 
0 & \sqrt{2}A\; & \;W+\Delta _{\omega }
\end{array}
\right) \text{,}  
\end{equation}
where $A\equiv \left| A\right| e^{i\phi }$ defines the real amplitude and
the phase of the electron-photon coupling. Diagonalization gives the
eigenenergies and eigenfunctions associated with Eq. (10). We get

\begin{eqnarray}
\nonumber
E^{3}-4WE^{2}+({5W}^{2}{-4}\left| {A}\right| {{^{2}}-{{\Delta }_{\omega }^{2}%
}})E+\hspace{-.05cm}
\\
2W(\Delta _{\omega }^{2}+2\left| A\right| ^{2}-W^{2})=0\text{,}\hspace{-.65cm}
\end{eqnarray}
as the eigenenergy equation. In resonance $\Delta _{\omega }\equiv 0$, hence we see that Eq. (11) has solutions: 
\begin{equation}
E_{{\small 0}}=W,\;\text{and}\;E_{{\small 1,2}}=
{\textstyle{1 \over 2}}
\left( 3W\pm \sqrt{16\left| A\right| ^{2}+W^{2}}\right) \text{.}
\end{equation}
The eigenenergies for the case of an off-resonance pulse of light can also be found 
analytically \cite{reina2}. We do not give the explicit expressions for
brevity. The eigenfunctions for the $N=2$ problem are given by:

\begin{eqnarray}
\left | E=E_{k}\right\rangle  =\Gamma _{k}\times\\
\hspace{-3.52cm}
\left(
\left| 0\right\rangle +
{\textstyle{E_{_{k}}+\Delta _{\omega }-W \over \sqrt{2}\left| A\right| }}
\left| 1\right\rangle -
{\textstyle{2\left| A\right| ^{2}+(2W-E_{_{k}})(E_{_{k}}+\Delta _{\omega }-W) \over 2\left| A\right| ^{2}}}
\left| 2\right\rangle\right) 
\text{,} \nonumber \hspace{-3.12cm}
\end{eqnarray}
where \\$\Gamma _{k}=\sqrt{2}\left| A\right| \left[ 4\left| A\right|
^{2}+\left( \Delta _{\omega }+W\right) (E_{_{k}}+\Delta _{\omega }-W)\right]
^{-\frac{1}{2}},$ with $E_{_{k}}$ given as above ($k=0,1,2$). The procedure
\mbox{ des-\/} cribed here enables us  to perform the calculation of the laser pulse length 
required for generating the searched entangled Bell states. 

In general, for any value of $N,$\ the total wave function associated with
the initial condition $\left| \Psi (t=0)\right\rangle =\left| \Psi
_{0}\right\rangle ${\it \ }can be expressed{\it \ }as $\left| \Psi
(t)\right\rangle _{\Lambda }=
\mathop{\displaystyle\sum}
_{k}C_{k}e^{-iE_{k}t}\left| \psi _{k}\right\rangle $,$\;$where $
\mbox{\ $H^{\prime }\left| \psi _{k}\right\rangle =E_{k}\left| \psi
_{k}\right\rangle  $,\/}\;$and $\left| \psi _{k}\right\rangle =
\mathop{\displaystyle\sum}
_{j}A_{kj}\left| M_{j}\right\rangle $.$\;$Here the normalization
coefficients $C_{k}$ \cite{Note3}\ depend on the chosen initial condition $%
\left| \Psi _{0}\right\rangle $. The matrix elements $A_{kj}$ must be
determined for each particular value of $N$, and $\left| M_{j}\right\rangle
\equiv \left| J,M_{j};q\right\rangle $ as indicated earlier in this
section. Hence, the total wave function $\left| \Psi (t)\right\rangle
_{\Lambda}$ can be written as: 
\begin{equation}
\left| \Psi (t)\right\rangle _{{\small \Lambda}}=
\mathop{\displaystyle\sum}
_{k}
\mathop{\displaystyle\sum}
\limits_{j}^{{}}
C_{k}A_{kj}e^{-iE_{k}t}\left| M_{j}\right\rangle \text{.}  
\end{equation}
For the case of $N=2$ QDs, it is a straightforward\mbox{ exer-\/} cise to compute the
explicit coefficients of Eq. (14) for both of the $J-$subspaces that span the Hilbert space
$SU(2)\otimes SU(2)$. Next we centre our attention on the discussion of finding the
conditions to produce the\mbox{ maxi-\/} mally entangled Bell states. To achieve this, we
project the state $ \left| \Psi _{Bell}\right\rangle \;$over the wave function\ given by
Eq. (14) obtaining the result:
\begin{equation}
\left\langle \Psi _{Bell}\right. \left| \Psi (t)\right\rangle _{{\small
\Lambda}}=
{\textstyle{1 \over \sqrt{2}}}
\mathop{\displaystyle\sum}
_{k}C_{k}\left(
A_{k1}+e^{i\varphi }A_{k3}\right)e^{-iE_{k}t} \text{.}  
\end{equation}
Under the unitary evolution of the Hamiltonian $H^{\prime }$,$\;$the density
of probability $\wp (Bell)\;$for finding the entangled Bell state in this
coupled QD system is proportional to $\left| \left( \left\langle 0\right|
+e^{i\varphi }\left\langle 2\right| \right) \left| \Psi (t)\right\rangle
_{\Lambda}\right| ^{2}$. More explicitly we find
\begin{equation}
\wp (Bell)=
{\textstyle{1 \over 2}}
\left| 
\mathop{\displaystyle\sum}
_{k}C_{k}\left(
A_{k1}+e^{i\varphi }A_{k3}\right)e^{-iE_{k}t}\right| ^{2}\text{.}  
\end{equation}
Results and discussion of the time evolution described by Eq. (16), for several
different combinations of the physical parameters in the model, are
discussed later.

\subsection{Coupling of $\bf N=3$ QDs and generation of GHZ states}

Here we address the problem of generation of entangled GHZ states of the form
\mbox{\ $\left| \Psi _{GHZ}\right\rangle =
{\textstyle{1 \over \sqrt{2}}}
(\left| 000\right\rangle +e^{i\varphi }\left| 111\right\rangle ),$\/} for any
$\varphi, $ in the proposed system of $3$ coupled QDs. In this case, the
Hilbert space $SU(2)^{\otimes 3}$ is spanned by the eight basis vectors
associated with the $3$ different $J-$subspaces. Without loss of generality,
consider the $J=\frac{3}{2}-$subspace as the only one optically active. We
introduce the notation \mbox{\ $\left| M_{1}\right\rangle \equiv \left|
3/2,-3/2\right\rangle \equiv \left| 0\right\rangle $,\/} \mbox{\ $\left|
M_{2}\right\rangle \equiv \left| 3/2,-1/2\right\rangle \equiv \left|
1\right\rangle $,\/} $\left| M_{3}\right\rangle \equiv \left|
3/2,1/2\right\rangle \equiv \left| 2\right\rangle $, and $\left|
M_{4}\right\rangle \equiv \left| 3/2,3/2\right\rangle \equiv \left|
3\right\rangle $ to denote the vacuum state, the single-exciton state, the
biexciton state and the triexciton state respectively. In the absence of
light, the energy levels of the system are given by $E_{0}\equiv E(3/2,-3/2)=
{\textstyle{3 \over 2}}
(W-\Delta _{\omega })$, $E_{1}\equiv E(3/2,-1/2)=
{\textstyle{1 \over 2}}
(7W-\Delta _{\omega })$, $E_{2}\equiv E(3/2,1/2)=
{\textstyle{1 \over 2}}
(7W+\Delta _{\omega })$, and $E_{3}\equiv E(3/2,3/2)=
{\textstyle{3 \over 2}}
(W+\Delta _{\omega })$. We note that, as indicated in the preceding section, the
energy separation $E_{3,0}\equiv E_{3}-E_{0}=3\Delta _{\omega }$ is unaffected
by the interdot interaction strength $W$. Now we consider the effect of the
pulse of light $\xi (t)$ over this system of 3 QDs in the
$J=\frac{3}{2}-$ subspace: the associated Hamiltonian is:

\begin{equation}
\widehat{H^{^{\prime }}}=
\left( 
\begin{array}{cccc}
\frac{3W}{2}-\frac{3\Delta _{\omega } }{2} & \sqrt{3}A^{\ast } & 0 & 0 \\ 
\sqrt{3}A & \frac{7W}{2}-\frac{\Delta _{\omega } }{2} & 2A^{\ast } & 0 \\ 
0 & 2A & \frac{7W}{2}+\frac{\Delta _{\omega } }{2} & \sqrt{3}A^{\ast } \\ 
0 & 0 & \sqrt{3}A & \frac{3W}{2}+\frac{3\Delta _{\omega } }{2} \\
\end{array} 
\right).
\end{equation}
Diagonalization leads us to the following 4{\it th} order equation: 

\begin{eqnarray}
\nonumber
\left( \left[ {\textstyle{3 \over 2}}
(W-\Delta _{\omega } )-E\right] \left[ {\textstyle{1 \over 2}}
(7W-\Delta _{\omega } )-E\right] -3\left| A\right| ^{2}\right)\times
\hspace{-0.5cm}
\\
\nonumber
\left( \left[ 
{\textstyle{3 \over 2}}(W+\Delta _{\omega } )-E\right] \left[ 
{\textstyle{1 \over 2}}(7W+\Delta _{\omega } )-E\right] -3\left| A\right|
^{2}\right) -
\hspace{-0.5cm} 
\\ 
\vspace{-0.45cm}
4\left| A\right| ^{2}\left( 
{\textstyle{3 \over 2}}(W-\Delta _{\omega } )-E\right) \left( 
{\textstyle{3 \over 2}}(W+\Delta _{\omega } )-E\right) =0\text{,}
\end{eqnarray}
which is non-trivial to solve analytically in the case of a pulse with 
arbitrary frequency $\omega $. However, if $\xi (t)$\ is applied at
resonance ($\Delta _{\omega }=0$), we get the following eigenenergies:  
\begin{eqnarray}
\nonumber
E_{0,1}={\textstyle{5 \over 2}}
W+\left| A\right| \pm \sqrt{\left( W+\left| A\right| \right) ^{2}+3\left|
A\right| ^{2}}\text{,}\;
\\
E_{2,3}={\textstyle{5 \over 2}}
W-\left| A\right| \pm \sqrt{\left( W-\left| A\right| \right) ^{2}+3\left|
A\right| ^{2}}\text{,}
\end{eqnarray}
with eigenvectors: 
\begin{eqnarray}
\nonumber
\left| E_{0,1}\right\rangle  &=&\eta _{_{0,1}}\times
\\
\hspace{-3.5cm}
\left[\left| 0\right\rangle+\left({\textstyle{E_{0,1}-\frac{3W}{2} \over
\sqrt{3}\left| A\right| }}\right) \left|1\right\rangle +\left(
{\textstyle{E_{0,1}-\frac{3W}{2} \over \sqrt{3}\left| A\right| }}
\right) \left|
2\right\rangle +\left| 3\right\rangle \right]
\text{,}\hspace{-3.5cm}
\end{eqnarray}
\begin{eqnarray}
\nonumber
\left| E_{2,3}\right\rangle  &=&\eta _{_{2,3}}\times
\\
\hspace{-3.5cm}
\left[
\left| 0\right\rangle
+\left(
{\textstyle{E_{2,3}-\frac{3W}{2} \over \sqrt{3}\left| A\right| }}
\right) 
\left|
1\right\rangle 
-\left(
{\textstyle{E_{2,3}-\frac{3W}{2} \over \sqrt{3}\left| A\right| }}
\right) 
\left|
2\right\rangle -\left| 3\right\rangle \right] \text{,} \hspace{-3.5cm}
\end{eqnarray}
where the coefficients $\;\eta _{i}=\frac{1}{\sqrt{2}}\left( 1+\frac{\left( E_{i}-\frac{3W}{2}
\right) ^{2}}{3\left| A\right| ^{2}}\right) ^{-\frac{1}{2}}$, with
$\;i=0$,...,$3$ are normalization constants. The associated
total wave function $\left| \Psi (t)\right\rangle _{\Lambda}\;$(Eq. (14))
depends on the chosen initial condition $\left| \Psi (t=0)\right\rangle \equiv
\left| \Psi _{0}\right\rangle $ and is a\mbox{ linear\/} combination of the
eigenfunctions (20) and (21). We have computed, in both rotating and
laboratory frames, the analytical expressions for $\left| \Psi
(t)\right\rangle _{\Lambda}$ for all of the initial conditions $\left| \Psi
_{0}\right\rangle =\left\{ \left| 0\right\rangle ,\left| 1\right\rangle
,\left| 2\right\rangle ,\left| 3\right\rangle \right\} $.$\;$As an example
of this procedure, we give the result for the zero-exciton state as the
initial state, i.e. \mbox{\ $\left| \Psi _{0}\right\rangle =\left|
M_{1}\right\rangle \equiv \left| 0\right\rangle $.\/} In this case, the wave
function $\left| \Psi (t)\right\rangle _{\Lambda}$ is spanned by the following
coefficients $C_{k}$ and $A_{kj}$: 
\begin{eqnarray}
\nonumber
C_{0}=\frac{E_{1}-\frac{3W}{2}}{2\eta _{0}\left( E_{1}-E_{0}\right) }\text{,}%
\;C_{1}=\frac{E_{0}-\frac{3W}{2}}{2\eta _{1}\left( E_{0}-E_{1}\right) }\text{%
,\ }
\\
C_{2}=\frac{E_{3}-\frac{3W}{2}}{2\eta _{2}\left( E_{3}-E_{2}\right) }%
\text{, }C_{3}=\frac{E_{2}-\frac{3W}{2}}{2\eta _{3}\left( E_{2}-E_{3}\right) 
}\text{;} 
\end{eqnarray}

\begin{equation}
\widehat{A}\equiv \left( 
\begin{array}{cccc}
\eta _{0} & \frac{(E_{0}-\frac{3}{2}W)e^{i\varphi }}{\sqrt{3}\left| A\right| }\eta _{0}
& \frac{(E_{0}-\frac{3}{2}W)e^{i2\varphi }}{\sqrt{3}\left| A\right| }\eta_{0} 
& e^{i3\varphi }\eta _{0} \\ 
\eta _{1} & \frac{(E_{1}-\frac{3}{2}W)e^{i\varphi }}{\sqrt{3}\left| A\right| }\eta _{1}
& \frac{(E_{1}-\frac{3}{2}W)e^{i2\varphi }}{\sqrt{3}\left| A\right| }\eta_{1} & e^{i3\varphi }\eta _{1} \\ 
\eta _{2} & \frac{(E_{2}-\frac{3}{2}W)e^{i\varphi }}{\sqrt{3}\left| A\right| }\eta _{2} & 
-\frac{(E_{2}-\frac{3}{2}W)e^{i2\varphi }}{\sqrt{3}\left| A\right| }\eta_{2} & -e^{i3\varphi }\eta _{2}
\\ 
\eta _{3} & \frac{(E_{3}-\frac{3}{2}W)e^{i\varphi }}{\sqrt{3}\left|
A\right| }\eta _{3} & -\frac{(E_{3}-\frac{3}{2}W)e^{i2\varphi }}{\sqrt{3}%
\left| A\right| }\eta _{3} & -e^{i3\varphi }\eta _{3}
\end{array}
\right) \text{.} 
\end{equation}

\bigskip

The density of probability $\wp (GHZ)\;$of finding the entangled GHZ state
between vacuum and triexciton states is given by: 
\begin{equation}
\wp (GHZ)=
{\textstyle{1 \over 2}}
\left| 
\mathop{\displaystyle\sum}
_{k}C_{k}\left(
A_{k1}+e^{i\varphi }A_{k4}\right) e^{-iE_{k}t}\right| ^{2}\text{.}  
\end{equation}
Analytical and numerical computations have been performed for all$\;$of the
initial conditions $\left| \Psi _{0}\right\rangle $ mentioned above. These
results enable us to obtain specific conditions for the realization of such
maximally entangled GHZ states starting from suitable $\varphi -$pulses
and experimental parameters $\epsilon ,W,$ and $A.$ Details are discussed in
section III.

The quantum dynamical problem given by Eq. (5) is easily expressed in terms
of the expansion coefficients $d_{M}(t)\;$of the wave function. As usual, we
write \mbox{\ $\left| \Psi (t)\right\rangle _{\Lambda }=
\mathop{\displaystyle\sum}
\nolimits_{M=-J}^{J}d_{M}(t)e^{-iE_{M}t}\left| M\right\rangle $,\/}$\;$
so the time dependent problem is reduced to finding the solutions of the
following set of $2J+1\;$linear differential equations:

\begin{eqnarray}
i\partial _{t}d_{{\small M}}(t) = 
\hspace{2.5cm}
\\
A\sqrt{J(J+1)-M(M-1)}\;e^{(E_{{\small M}}-E_{{\small M-1}}-\omega
)it}d_{{\small M-1}}(t)+ 
\nonumber 
\hspace{-0.89cm}
\\
A^{\ast }\sqrt{J(J+1)-M(M+1)}\;e^{(E_{{\small M}}-E_{{\small M+1}}+
\omega )it}d_{{\small M+1}}(t)\text{,} \hspace{-0.7cm}
\nonumber
\end{eqnarray}
where $E_{M}=E(J,M)+\omega $.$\;$More explicitly, $E_{M,M-1}\equiv
E_{M}-E_{M-1}=\epsilon +W\left[ 1-2M\right] $, $E_{M,M+1}=W\left[
1+2M\right] -\epsilon $, and the problem given by Eq. (25) can be expressed
in terms of reduced units as follows: 

\begin{eqnarray}
i\partial _{\tau }f_{{\small M}}(\tau ) =
\hspace{2.5cm}
\\
\lambda \sqrt{J(J+1)-M(M-1)}
\;e^{[1+\mu (1-2M)-\nu ]i\tau }f_{{\small M-1}}(\tau )+  
\nonumber 
\hspace{-0.65cm}
\\
\lambda ^{\ast }\sqrt{J(J+1)-M(M+1)}\;e^{-[1-\mu (1+2M)-\nu ]i\tau }f_{
{\small M+1}}(\tau )\text{,}  \hspace{-0.7cm}
\nonumber
\end{eqnarray}
with the dimensionless parameters $\lambda =\frac{A}{\epsilon },$ $\mu =%
\frac{W}{\epsilon },$\ $\nu =\frac{\omega }{\epsilon },\ \tau ={%
\epsilon }t,$\ and $d_{M}(t)=f_{M}(\tau )\;$. The
set of Eqs. (26) gives the dynamics for any number of QDs and $(J,M)$ values. 
Numerical solutions were found
by varying the parameters $ \lambda ,\mu $, and $\nu .\;$Comparison between these
numerical solutions and the analytical ones yields
excellent agreement for the generation of both Bell and GHZ states, starting
from suitable initial conditions $\left| \Psi _{0}\right\rangle .$ Results
will be addressed in the next section.

\section{Results and Discussion}

In this section we discuss the main results obtained from the computation
of both analytical and numerical solutions for the unitary evolution 
described in the preceding section. Figure 1 shows the
probability density for finding the entangled Bell state ($N=2$) between vacuum and
biexciton states given by Eq. (16) as a function of time for the initial
condition $\left| \Psi _{0}\right\rangle =\left| 0\right\rangle $. As seen from Fig. 1, 
selective pulses of length $\tau _{_{B}}$ can be used to
create maximally entangled Bell states in the system of two coupled QDs. The energy $W$ is kept fixed while
the amplitude of the radiation pulse $A$ is varied. The results indicate that the
time $\tau _{_{B}}$ is increased with diminishing incident field strength $A$. 
As an example, we consider wide-gap semiconductor
QDs, like ZnSe based QDs, with band gap $\epsilon =2.8\;e$V and resonant
optical frequency $\omega =4.3\times 10^{15}$s$^{-1}$. For a $0$ or $2\pi
-$pulse, $W=0.1\;$and$\;A=
{\textstyle{1 \over 25}}%
$, Fig. 1(a) shows that the generation of
the state\thinspace $
{\textstyle{1 \over \sqrt{2}}}
\left( \left| 0\right\rangle +\left| 2\right\rangle \right) $ requires a
pulse of length $\tau _{_{B}}=7.7\times 10^{-15}$s. By changing the value of
the\mbox{ para-\/} meter $A$ (see Figs. 1(b), (c) and (d)), we can modify
the length $\tau _{_{B}}$ of this Bell pulse (Fig.
1 covers the interval $10^{-11}$s $<\tau _{_{B}}<10^{-15}$s$)$. Another method
for manipulating the length $\tau _{_{B}}$ is shown in Fig. 2. Here we vary $W$ for fixed $A=10^{-3}$. Experimentally, this
variation of $W$ can be tailored by changing the
interdot distance. In this case, the analysis shows that for a fixed value of
$A$ the length $\tau _{_{B}}$ decreases with decreasing interaction strength $W$.

The same investigation of parameter dependance was performed for
the case of the entangled GHZ state ($N=3$) between vacuum and triexciton
states. We calculate the probability given by Eq.
(24) as a function of time, starting with the initial condition $\left| \Psi
_{0}\right\rangle =\left| 0\right\rangle $. Figure 3 shows the selective
pulses used to create such maximally entangled GHZ states in the system of
three coupled QDs. For example, in the case of Fig. 3(a), the\mbox{ gene-\/} ration of
the GHZ state $
{\textstyle{1 \over \sqrt{2}}}
\left( \left| 0\right\rangle +\left| 3\right\rangle \right) $ requires a
time $\tau _{_{GHZ}}=1.3\times 10^{-14}$s. Figures 3(b), (c), and (d) explore
several different ranges for the $\tau _{_{GHZ}}-$pulses required in the
generation of such GHZ states. For fixed $W$, the time 
$\tau _{_{GHZ}}$ increases with decreasing incident field
strength $A.$ In contrast, for fixed $A,$ the length $
\tau _{_{GHZ}}$ decreases with decreasing interdot interaction
strength.

The above results are not restricted to ZnSe-based QDs: by employing
semiconductors of different bandgap $\epsilon $ (e.g. GaAs), other
regions of parameter space can be explored. We have studied the time
evolution of the system of QDs for several different values of the phase $
\varphi $. These give similar qualitative results to the ones discussed previously. Here we only
include the $0$ or $2\pi -$pulse results since these are ones corresponding to
the discussion in Sec. IV. The relevant experimental
conditions as well as the required coherent control to realize
the above combinations of parameters,
are compatible with those\mbox{ demonstrated\/} in Ref.\cite{Bonadeo1}. We point
out that the procedure described in this paper is valid for any
value of the phase constant $\varphi$, in contrast to Ref.\cite{Quiroga} 
where analytic results were derived for the
particular case $\varphi=\frac{\pi}{2}.$ The generation of maximally
entangled states in this paper has considered the\mbox{ experi-\/} mental
situation of global laser pulses only; however, by using near-field optical
spectroscopy \cite{ChavezPirson}, individual QDs from an ensemble can be addressed by using local pulses,
a feature that can be exploited to generate entangled states with different
symmetries, such as the antisymmetric state ${\textstyle{1 \over \sqrt{2}}}
(\left| 01\right\rangle -\left| 10\right\rangle).\;$ We will hence
be able to generate the complete Bell basis consisting of four mutually orthogonal states for
the 2 qubits, all of which are maximally entangled, i.e. the set of states
${\textstyle{1 \over \sqrt{2}}}\{(\left| 00\right\rangle +\left|
11\right\rangle),\; (\left| 00\right\rangle -\left| 11\right\rangle),\; (\left|
01\right\rangle +\left| 10\right\rangle),\; (\left|
01\right\rangle -\left| 10\right\rangle)\} $.$\;$ From a general point of
view, this basis is of fundamental relevance for quantum
information processing. We stress that the optical generation of
excitonic entangled states in coupled QDs given here could be exploited in
solid state devices to perform quantum protocols, as recently proposed in Ref.
\cite{Reina} for teleporting an excitonic state in a coupled QD
system.

\section{Quantum Information Processing in Coupled Dots}

To perform quantum computation operations, we can use an initial pure state followed by 
a series of transformations on this state using unitary
operations. \mbox{ Another\/} possibility is to use an
initial mixed state, providing the decoherence time is sufficiently long 
\cite{NMR}. In order to implement such quantum operations we need two
elements: the {\it %
Hadamard} transformation and the quantum CNOT gate.
In the orthonormal computation basis of single qubits $\left\{ \left|
0\right\rangle ,\left| 1\right\rangle \right\} $, the CNOT gate acts on two
qubits $\left| \varphi _{i}\right\rangle $ and $\left| \varphi _{j}\right\rangle $ 
simultaneously
as follows: CNOT$_{ij}(\left| \varphi _{i}\right\rangle \left| \varphi
_{j}\right\rangle )\mapsto \left| \varphi _{i}\right\rangle \left| \varphi
_{i}\oplus \varphi _{j}\right\rangle.\; $Here $\oplus$ denotes {\it 
addition\ modulo}$\;2$, and the indices $i$ and $j$ refer to the control bit
and the target bit respectively. The Hadamard transformation $H^{T}\;$acts
only on single qubits by performing the rotations: $H^{T}(\left|
0\right\rangle )\mapsto \frac{1}{\sqrt{2}}\left( \left| 0\right\rangle
+\left| 1\right\rangle \right), \;$and $H^{T}(\left| 1\right\rangle )\mapsto 
\frac{1}{\sqrt{2}}\left( \left| 0\right\rangle -\left| 1\right\rangle
\right) $. In our scheme, $\left| 0\right\rangle \;$represents the
vacuum state for excitons while $\left| 1\right\rangle \;$represents a
single exciton. Experimental demonstration of single qubit rotations (and hence the Hadamard transformation)
in the case of individual excitons confined to QDs should now be possible, as a result of a 
recent experiment reporting direct observation of excitonic Rabi oscillations
in semiconductor quantum wells \cite{Schulzgen}. Despite the fact that Rabi
Flopping in QDs is still under intensive experimental study, results
given in Ref.\cite{Schulzgen} lead us to believe that we are not too far away from
the experimental observation of such excitonic Rabi oscillations in QDs and
hence the demonstration of single qubit operations.

The adequate preparation, computation and readout of information, in addition to the
coherent coupling of the qubits to the environment, are compulsory
steps for the successful construction of a universal quantum computer. We briefly
review these requirements and their relationship with the model proposed here: (1) {\it A very well defined Hilbert space}: We must have an
adequate control over the Hilbert space of qubits. In our scheme, the
orthonormal computation basis of single qubits is represented by the vacuum
state ($\left| 0\right\rangle)\;$and the single state of excitons ($\left|
1\right\rangle ).\;$ (2) {\it Initializing the computer}: Before commencing any quantum
computation task, we need a rapid relaxation of our
qubits to their ground state, i.e. zero excitons per dot. In our case, numerical
values indicate that this state is easily achieved by turning the laser off and waiting
for a few femtoseconds. (3) {\it Inputting initial data\ and readout}: As pure and entangled
states with different symmetries can be obtained using the experimental
techniques described in Ref.\cite{Bonadeo1,Bonadeo2,ChavezPirson}, we would have the
ability to manipulate the input of the quantum state of the QD system. Hence,
we would have the experimental possibility to control the optical excitation and to
detect individual QD signals from an entire dot ensemble, thereby facilitating
individual qubit control for the readout. In fact, nanoprobing enables us to
measure directly the excitonic and biexcitonic luminescence from single QDs \cite{ChavezPirson}. (4) 
{\it Universal set of gate operations}: We need to be able to perform single 
qubit rotations and two qubit gates. We stress that the generation of the
maximally entangled states shown in Figs. $1-3$ corresponds to the physical
realization of a Hadamard transformation followed by a CNOT operation in the Bell
case, and two CNOT operations in the GHZ case. As to the practical semiconductor
nanostructure implementation, this set of gate operations is in a
preliminary stage of investigation and demands intensive
experimental study. 
(5) {\it Decoherence and the coupling to the
environment}: By taking into account \mbox{\ decoherence\/} mechanisms
(exciton-acoustic-phonon type) on the process of generation of the entangled
states discussed here, a recent work by Rodr\'{\i}guez {\it et al.}
\cite{Rodriguez} has shown that this generation is preserved over a reasonable
parameter window hence giving the possibility of performing the unitary
transformations required for quantum computing before decoherence of the excitonic
states takes place. 

\section{Conclusions}

In summary, we have solved both analytically and numerically the quantum mechanical
equation-of-motion for excitons in two and three coupled QD systems driven by
classical pulses of light. By doing this, we have been able to provide a mechanism for
preparing maximally entangled Bell and GHZ states via excitons in
optically-driven QDs, exploiting current levels of coherent optical control
such as the ones demonstrated using ultra-fast spectroscopy
\cite{Bonadeo1,Bonadeo2} and near-field optical spectroscopy
\cite{ChavezPirson}. This mechanism enables us to generate single
qubit rotations, such as the Hadamard one, and quantum CNOT gates. In
particular, the procedure presented here leads us to the generation of the
whole Bell basis, a fact that can be exploited in the process of quantum
teleportation of excitonic states in systems of coupled QDs \cite{Reina}.
Furthermore, by taking into account the main decoherence mechanisms, e.g. 
exciton-acoustic-phonon, we find that this optical generation of quantum
entanglement is preserved over a reasonable parameter window. This leads 
to the possibility of performing several hundred quantum
computation operations before decoherence of these excitonic states takes
place.

\section{Acknowledgments}
J.H.R. and L.Q. acknowledge the support of COLCIENCIAS. J.H.R. thanks D.J.T. 
Leonard for helpful discussions, and the
hospitality of the ESF-QIT programme meeting 1999 in Cambridge, where part 
of this work was performed.

{\small
FIG. 1. Generation of the Bell State $
{\textstyle{1 \over \sqrt{2}}}
(\left| 00\right\rangle +\left| 11\right\rangle )$.\ These pulses correspond
to the realization of the Hadamard gate followed by a quantum CNOT
gate. $W=0.1$,$\;\varphi =0,\;$
and (a)$\; A={\textstyle {1 \over 25}}$, (b)$\; A={\textstyle {1 \over
50}}$, (c)$\;A=10^{-2}$, and (d)$\;A=10^{-3}$. In the Figures $1-3$, $\left| \Psi
(t)\right\rangle$ denotes the total wavefunction of the system at time $t$ in both laboratory
(solid curves) and rotating frames (dashed curves). The energy is in units of the
band gap $\epsilon $, and $\left| \Psi _{0}\right\rangle =\left|
0\right\rangle $ in all of the figures. \bigskip 

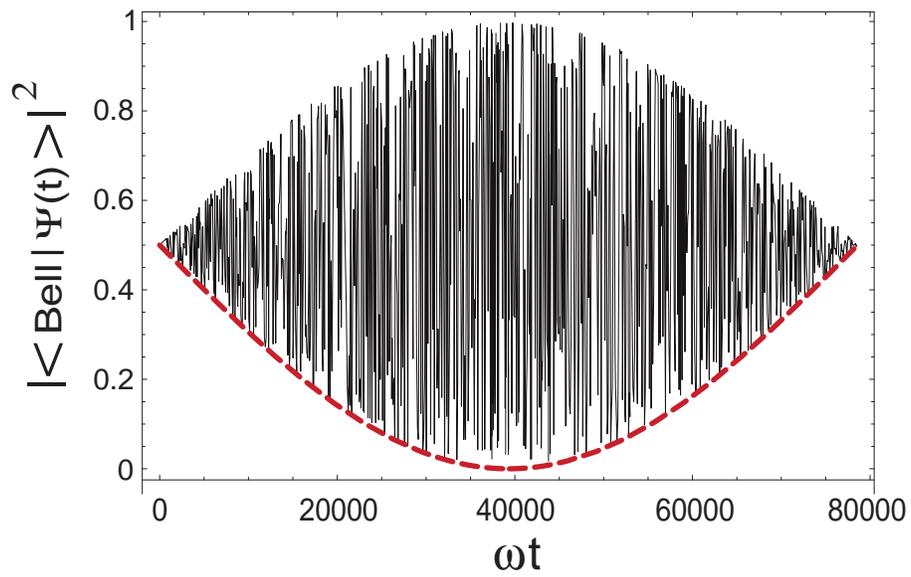
FIG. 2. Generation of the Bell state ${\textstyle{1 \over \sqrt{2}}}
(\left| 00\right\rangle +\left| 11\right\rangle )$.\ \mbox{\ $A=10^{-3}$\/},$\;\varphi
=0,\;$and (a) $W=0.1$, (b) $W=0.05$, and (c) $W=10^{-2}$.\bigskip

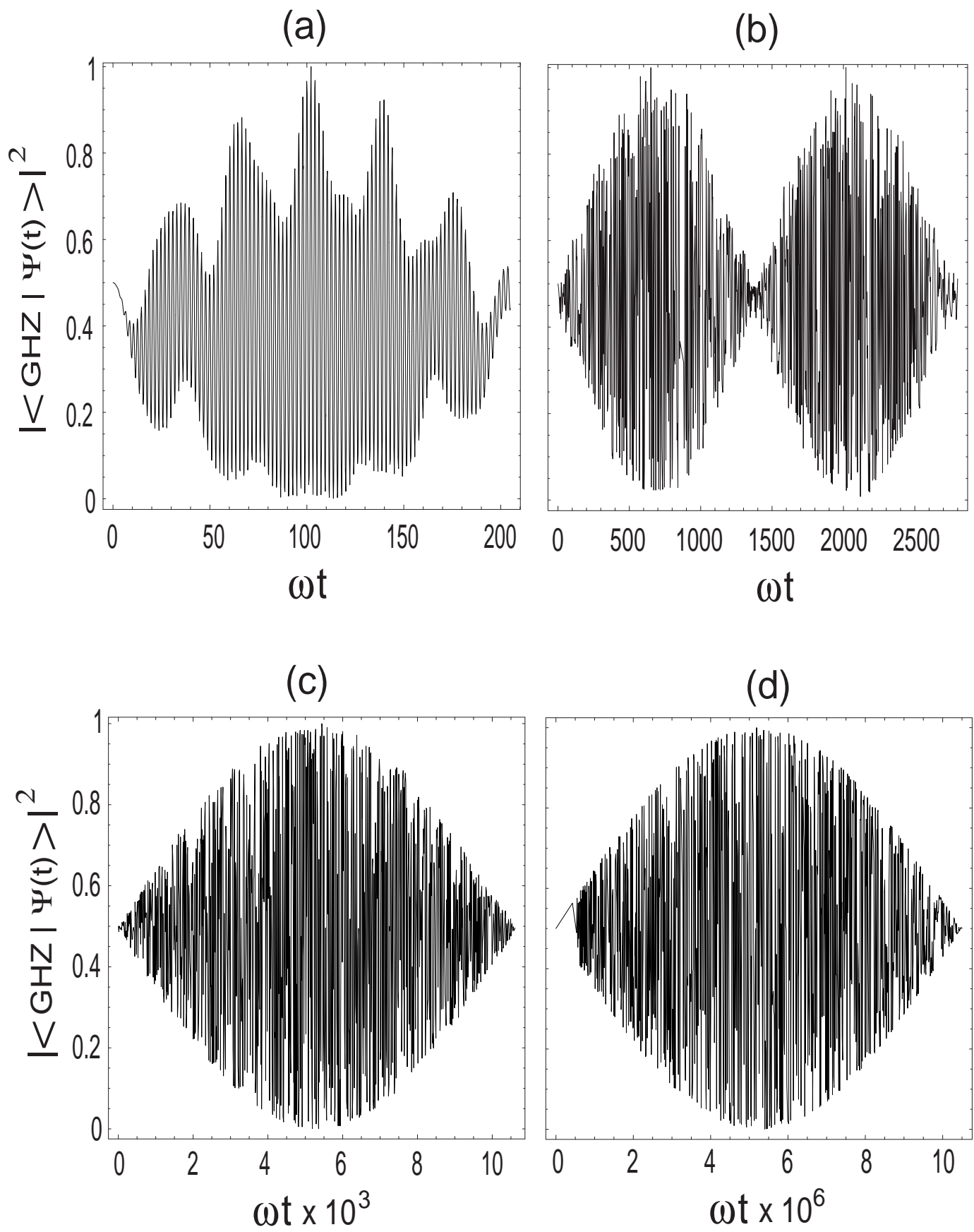
FIG. 3. Generation of the GHZ state ${\textstyle{1 \over \sqrt{2}}}
(\left| 000\right\rangle +\left| 111\right\rangle )$.\ These pulses correspond
to the realization of the Hadamard gate followed by two quantum CNOT
gates. $W=0.1$,$\;\varphi =0,\;$ and 
(a)$\; A={\textstyle {1 \over 25}}$, (b)$\; A={\textstyle {1 \over
50}}$, (c)$\;A=10^{-2}$, and (d)$\;A=10^{-3}$.} 

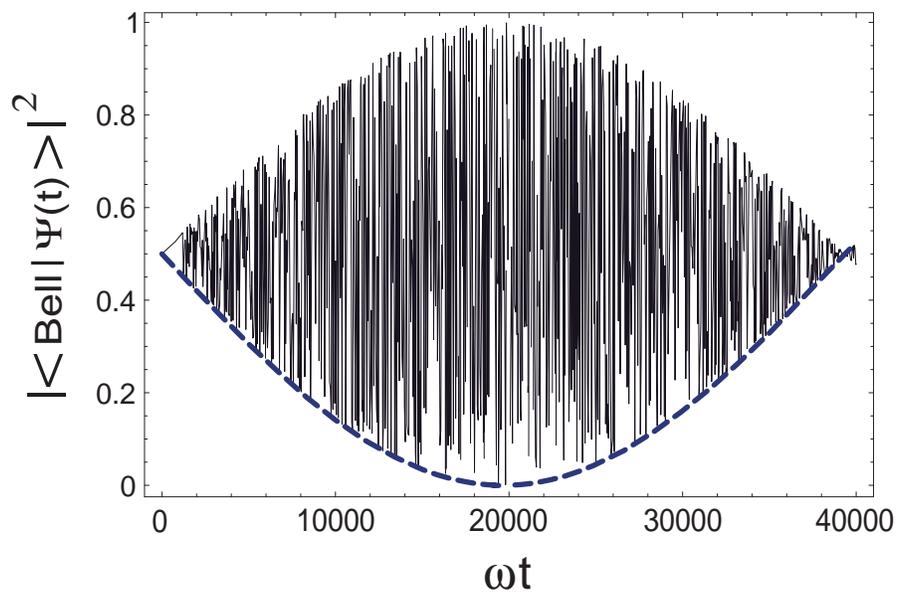
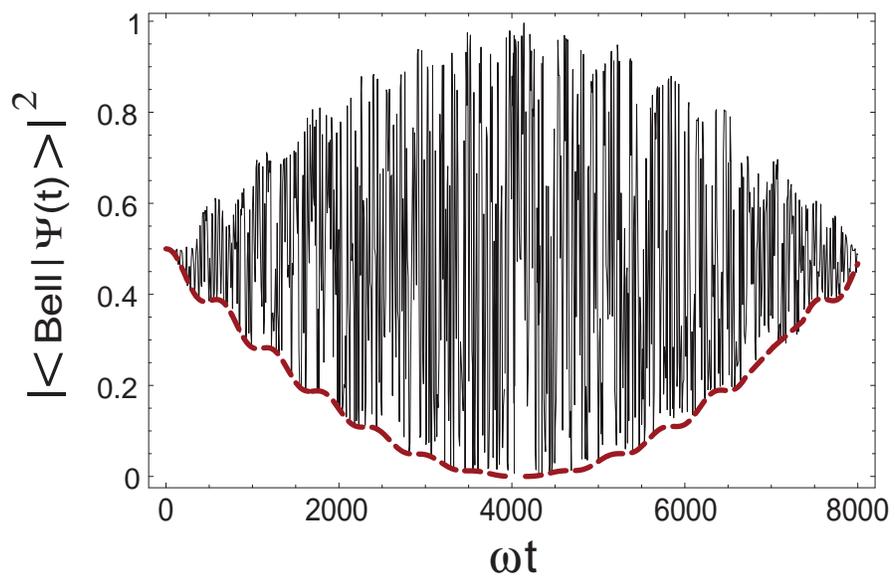
\end{document}